\documentclass[twocolumn,pra,showpacs]{revtex4}
\usepackage{amsmath,psfrag,graphicx,times,mathptmx}
\def\ket#1{|#1\rangle}
\def\bra#1{\langle #1|}
\def\bracket#1{\langle #1 \rangle}

\def\sub#1{_{\scriptsize \mbox{#1}}}

\begin{document}
\preprint{pra1.0}
\title{Generation of photon pairs using polarization-dependent
two-photon absorption}
\author{T. Nakanishi}
\email{t-naka@giga.kuee.kyoto-u.ac.jp}
\author{K. Sugiyama}
\author{M. Kitano}
\affiliation{Department of Electronic Science and Engineering,
Kyoto University\\
Kyoto 606-8501, Japan}
\date{\today}

\begin{abstract}
We propose a new method for generating photon pairs 
from coherent light using
polarization-dependent two-photon absorption.
We study the photon statistics of two orthogonally polarized modes
by solving a master equation,
and show that when we prepare a coherent state in one polarization mode,
photon pairs are created in the other mode.
The photon pairs have the same frequency as
that of the incident light.
\end{abstract}

\maketitle

\section{Introduction}\label{Sec:Intro}

The photon pair shows a variety of nonclassical phenomena
such as non-local quantum correlation \cite{Aspect:PRL82},
two-photon interference \cite{Mandel:RMP99}, and so on.
There are interesting applications which include 
quantum teleportation \cite{Bennett:PRL93},
quantum cryptography \cite{Ekert:PRL92},
and quantum computing \cite{Bouwmeester:BOOK00}.
Usually photon pairs are generated with
two-photon cascade emission \cite{Clauser:PRD73}
or spontaneous parametric
down-conversion \cite{Burnham:PRL70,Friberg:PRL85}.
In this paper we propose a new method to produce photon pairs
using polarization-dependent two-photon absorption from coherent light.

Two-photon absorption is one of the second-order 
nonlinear optical effects and many investigations 
have been performed \cite{Georgiades:PRA97,Ezaki:PRL99}.
Some researchers have shown that coherent light is changed into 
sub-Poissonian light with nonclassical characteristics after
the interaction with the two-photon absorbers
\cite{Loudon:OC84,Tornau:OC74,Loudon:BOOK}.
The rate of two-photon transition is proportional to the second-order
correlation function of the photon field \cite{Gilles:PRA93}.
The shorter the time interval between two photons
incident on a two-photon absorber,
the higher the probability of the absorption.
If the initial state of the field is a coherent state
whose photons are randomly distributed in time,
pairs of photons which arrive at the absorber
accidentally within a very short time period
are likely to be absorbed.
Hence the unabsorbed photons are apart from each other,
and exhibit antibunched photon statistics 
with low second-order coherence \cite{Krasinski:OC76}.

The above investigations deal with the polarization-independent
two-photon transition.
But there exists two-photon transition with a polarization selection
rule which allows only the transition by two photons 
whose polarizations are orthogonal each other.
When the polarization-dependent two-photon absorber
interacts with the coherent light that includes quantum superposition
of an absorbable and an unabsorbable two-photon states,
the time-correlated photon pairs 
in the unabsorbable state and isolated photons survive
after the absorption.
The polarization selective absorption accompanies
a change of the polarization state for the photon pairs,
and therefore, photon pairs with polarization orthogonal
to the original polarization are created. 
One can single out the photon pairs using a linear polarizer.
We will give the details in Sec.~\ref{Sec:PDTPA} and
analyze the time evolution of the field
in two orthogonal linear polarization modes
using a master equation in Sec.~\ref{Sec:Master_Eq}.

While photon pairs created by parametric down-conversion
have half the frequency of the pump laser,
the frequency of photon pairs generated by our method 
is the same as that of the coherent light initially prepared. 
Hence, choosing appropriate two-photon transitions,
one could produce photon pairs in blue or much higher frequency region.

\section{Polarization-dependent two-photon absorption}
\label{Sec:PDTPA}

\begin{figure}[b]
\begin{center}
 \psfrag{HH}[c][c]{$\ket{2,0}$}
 \psfrag{VV}[c][c]{$\ket{0,2}$}
 \psfrag{HV}[c][c]{$\ket{\Psi\sub{a}}$}
 \psfrag{ini}[c][c]{$\ket{\Psi\sub{i}}$}
 \psfrag{final}[c][c]{$\ket{\Psi\sub{f}}$}
 \psfrag{VH}[c][c]{$\ket{\Psi\sub{u}}$}
  \includegraphics[scale=.8]{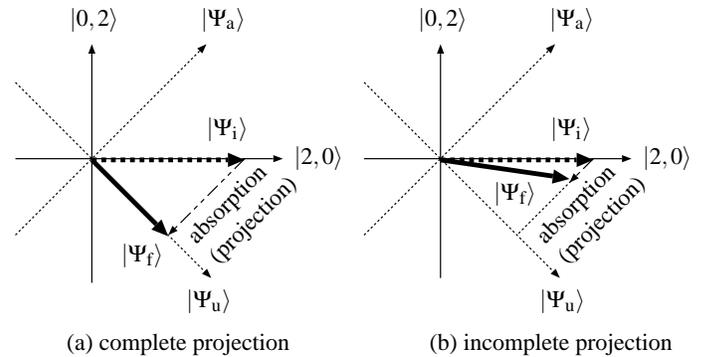}
 \caption{Projection of the state 
 by polarization-dependent two-photon absorption.
 The unabsorbable state $\ket{\Psi\sub{u}}$ is defined as 
 $(\ket{2,0}-\ket{0,2}) / \sqrt{2}$.}
 \label{Fig1}
\end{center}
\end{figure}

The two-photon transition between two $L=0$ states
obeys a polarization selection rule
which only allows the absorption of oppositely circular-polarized photons
\cite{Bonin:JOSAB84,Melikechi:JOSAB84}.
The selection rule is originated from the conservation of angular
momentum of the atom and the two photons, i.~e.~,
the photons must have zero angular momentum
if the initial state and the final state of the atom have 
zero angular momentum.
If one photon is right circularly polarized,
the other must be left circularly polarized.

In this section,
we will give simple explanations how photon pairs are
generated by the  polarization-dependent two-photon absorption.
We introduce the two-photon state
$\ket{\Psi\sub{a}}=\ket{1,1}\sub{RL}$,
where $\ket{r, l}\sub{RL}$ denotes the state with $r$
right circularly polarized photons and $l$ left circularly polarized photons.
We can  represent $\ket{\Psi\sub{a}}$
in the linear polarization basis as
\begin{align}
 \ket{\Psi\sub{a}}
 = \frac{\ket{2, 0}\sub{HV}+\ket{0, 2}\sub{HV}}{\sqrt{2}}, \label{Eq:Cond_TPA}
\end{align}
where $\ket{h, v}\sub{HV}$
is the state with $h$ photons in the horizontal 
polarization mode and $v$ photons in the vertical polarization mode
\cite{Burlakov:PRA99}.

First, we consider a simple case
where photon field is prepared in
$\ket{\Psi\sub{i}}=\ket{2, 0}\sub{HV}$
and interacts with the polarization-dependent two-photon absorber.
If the interaction time is long enough,
the state of the light is projected into the state
which is orthogonal to the state $\ket{\Psi\sub{a}}$,
as illustrated in Fig.~\ref{Fig1}(a).
Then the conditioned state for no absorption is represented by
\begin{align}
 \ket{\Psi\sub{f}} = (1-\ket{\Psi\sub{a}}\bra{\Psi\sub{a}}) \ket{2, 0}\sub{HV}
 = \frac{\ket{2, 0}\sub{HV}-\ket{0, 2}\sub{HV}}{2}, \label{Eq:superposition}
\end{align}
where the decrease of the norm is the result of the absorption.
Note that there is a probability
that one finds the vertically polarized two-photon state $\ket{0, 2}\sub{HV}$,
which did not exist before the interaction.
A similar process for one photon 
is occurred when a horizontally polarized photon 
passes through a linear polarizer aligned at 45 degrees.
The polarization is projected along the orientation of the polarizer,
and the vertical polarization component appears \cite{Sakurai:BOOK}.

In the case of incomplete absorption, as shown in Fig.~\ref{Fig1}(b),
the conditioned state for no absorption is given by
\begin{align}
 \ket{\Psi\sub{f}} =
 (1 -  \epsilon \ket{\Psi\sub{a}}\bra{\Psi\sub{a}}) \ket{2, 0}\sub{HV}
 = \left( 1 - \frac{\epsilon}{2} \right) \ket{2, 0}\sub{HV}
 - \frac{\epsilon}{2} \ket{0, 2}\sub{HV},
  \label{Eq:imcomplete_absorption}
\end{align}
where $\epsilon (>0)$ represents absorption.
For $\epsilon \ll 1$, the probability for $\ket{2, 0}\sub{HV}$
is reduced by $\epsilon$ and that for $\ket{0, 2}\sub{HV}$
is increased by $\epsilon^2/4$.
This change in polarization creates the vertically polarized
two-photon state, which did not exist initially.

\begin{figure}[b]
\begin{center}
 \includegraphics[scale=0.5,angle=-90]{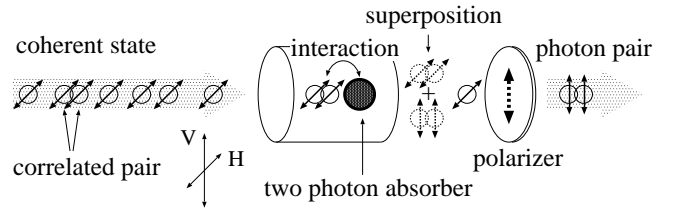}
 \caption{Procedure of generating photon pairs with two-photon absorber.}
 \label{Fig2}
\end{center}
\end{figure}

Next, we consider the case where
the initial state of light is a horizontally polarized 
coherent state as shown in Fig.~\ref{Fig2}.
Because the rate of the two-photon absorption is proportional to
the second-order correlation function of the field,
pairs within a extremely short interval are absorbed selectively
from the randomly distributed photons.
Therefore, only the time-correlated pairs
are changed into the superposition as
Eq.~(\ref{Eq:imcomplete_absorption}) by the absorption.
The pairs with horizontal polarization are reduced,
while vertically polarized photon pairs emerge
with probability $\epsilon^2/4$ for all pairs. 
All uncorrelated and isolated photons remain horizontally polarized.

The horizontally polarized beam is evolved from 
the coherent state into antibunched light due to the pair-wise absorption.
On the other hand, the vertical polarization mode is a vacuum at first,
and evolved into the bunching state by pair production of photons.
As shown in Fig.~\ref{Fig2},
filtering out horizontally polarized photons with a polarizer,
one can get photon pairs.

In the next section, we will analyze the photon statistics of
the horizontal and vertical polarization mode 
for quantitative evaluations.

\section{Master Equation Approach}
\label{Sec:Master_Eq}

We examine the time evolution of the field under the interaction with
the polarization-dependent two-photon absorbers
by solving a master equation.
Schauer  \cite{Schauer:PRL93} investigated the steady state solutions
of the master equation for the interaction between the 
two-photon absorbers and the photon field.
On the other hand we take a perturbative approach
under the condition of short interaction time.

We assume that the one-photon transitions to intermediate levels
are off resonant enough to be negligible compared with the 
two-photon transition.
We introduce a two-photon annihilation operator $O$:
\begin{align}
 O  = a\sub{R} a\sub{L}  = \frac{a\sub{H}^2+a\sub{V}^2}{2}, \label{Eq:O}
\end{align}
where $a\sub{R}$, $a\sub{L}$, $a\sub{H}$, and $a\sub{V}$
are annihilation operators for the right circularly,
left circularly, horizontally, and vertically polarized photons, respectively.
The density matrix of the light field, $\rho$, satisfies the following
master equation \cite{Tornau:OC74,Schauer:PRL93}:
\begin{align}
  \frac{d \rho}{d \tau} = 2 O \rho O^\dag 
  -  O^\dag  O \rho - \rho O^\dag  O, \label{Eq:masterEq}
\end{align}
where the time $\tau$ is normalized with respect to
 the two-photon transition rate $\kappa$ as $\tau=\kappa t$.
Equation (\ref{Eq:masterEq}) yields the equation of motion for an
operator $A$ as
\begin{align}
 \frac{d}{d \tau} \bracket{A}
 = 2 \bracket{O^\dag A  O} - \bracket{A O^\dag  O} - \bracket{ O^\dag O  A }. 
 \label{Eq:A}
\end{align}

We will calculate the evolution of the light field
when the initial state of the light is an intense
coherent state $\ket{\alpha}\sub{H}$
for the horizontal polarization mode and the vacuum state $\ket{0}\sub{V}$
for the vertical polarization mode, i.~e. 
$\ket{\alpha, 0}\sub{HV} \equiv \exp( \alpha a\sub{H}^\dag
 - \alpha^* a\sub{H} ) \ket{0}$.
As mentioned before, we assume that 
the normalized interaction time $T$ is much less than unity.
This is reasonable for general experimental realizations
without special equipments such as a high-{\it Q} cavity
because of very small two-photon transition rate.

\subsection{Horizontal polarization mode}

With Eq.~(\ref{Eq:A}), the rate of change for the number of
horizontally polarized photons can be easily obtained as
\begin{align}
 \frac{d}{d \tau}\bracket{a^\dag\sub{H} a\sub{H}}
 = - \bracket{a^{\dag 2}\sub{H} a^2\sub{H}} 
 - \frac{1}{2}\bracket{a^{\dag 2}\sub{H} a^2\sub{V}}
 - \frac{1}{2}\bracket{a^2\sub{H} a^{\dag 2}\sub{V}}. \label{Eq:num_a} 
\end{align}
The interaction time is so small
that the horizontally polarized beam still remains to be very intense
and the vertical polarization is nearly in the vacuum mode 
after the interaction.
Then, the first term of the right hand side of Eq.~(\ref{Eq:num_a})
is dominant:
\begin{align}
 \frac{d}{d \tau}\bracket{a^\dag\sub{H} a\sub{H}}
 \simeq -\bracket{a^{\dag 2}\sub{H} a^2\sub{H}}. \label{Eq:num_a2} 
\end{align}
The reduction rate of the photon number 
is proportional to the second-order correlation function of the field.

\begin{figure}[t]
\begin{center}
 \psfrag{x1}[c][c]{\footnotesize $X_{\rm H}(0)$}
 \psfrag{x2}[c][c]{\footnotesize $X_{\rm H}(\pi/2)$}
 \psfrag{y1}[c][c]{\footnotesize $X_{\rm V}(0)$}
 \psfrag{y2}[c][c]{\footnotesize $X_{\rm V}(\pi/2)$}
 \includegraphics[scale=0.5,angle=-90]{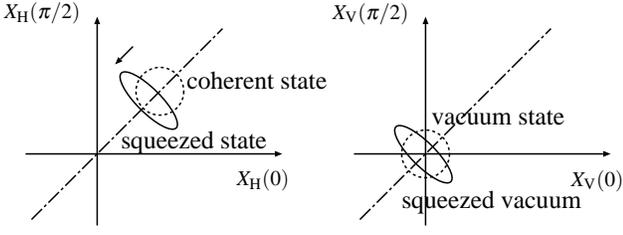}
 \caption{Time evolution in the phase space
 (a) for the horizontal polarization
 mode and (b) for the vertical polarization mode.
 The dotted lines represent initial states and the solid lines represent
 the state after the interaction with the absorber.}
 \label{Fig3}
\end{center}
\end{figure}

From Eq.~(\ref{Eq:num_a2}), the mean photon number after the interaction 
can be approximated as
\begin{align}
 \bracket{a^\dag\sub{H} a\sub{H}} 
 \simeq |\alpha|^2 - |\alpha|^4 T. \label{Eq:num_a3} 
\end{align}
Similarly, other resulting expectation values are calculated as
\begin{align}
 &\bracket{a\sub{H}} \simeq \alpha - \frac{|\alpha|^2 \alpha  T}{2} , 
 \nonumber \\
 &\bracket{a^2\sub{H}} \simeq \alpha^2 - |\alpha|^2 \alpha^2 T
 - \frac{\alpha^2 T}{2} , \label{Eq:a}\\
 &\bracket{a^{\dag 2}\sub{H} a^2\sub{H}} \simeq |\alpha|^4 - 2 |\alpha|^6 T 
 -  |\alpha|^4 T. \nonumber
\end{align}
We introduce an Hermitian operator 
$X\sub{H}(\phi)=a\sub{H} e^{- i \phi} + a^\dag\sub{H} e^{i \phi}$,
and calculate its dispersion $\bracket{(\Delta X\sub{H}(\phi))^2}$
from Eqs.~(\ref{Eq:num_a3}) and (\ref{Eq:a}):
\begin{align}
 \bracket{(\Delta X\sub{H}(\phi))^2} \simeq 1 - |\alpha|^2 T \cos 2({\rm arg}\,
 \alpha -\phi), \label{Eq:Xa}
\end{align}
which shows that the dispersions of the quadrature amplitudes
$X\sub{H}({\rm arg}\,\alpha)$ and $X\sub{H}({\rm
arg}\,\alpha+\frac{\pi}{2})$ satisfy $\bracket{(\Delta X\sub{H}({\rm
arg}\,\alpha))^2} < 1 < \bracket{(\Delta X\sub{H}({\rm arg}\,\alpha+
\frac{\pi}{2}))^2}$.
We see that the mode of the horizontal polarization 
is evolved into the squeezed state,
where the dispersion for $\phi={\rm arg}\,\alpha$ is smaller than unity.
As in Fig.~\ref{Fig3} (a),
the fluctuation in the direction of the excitation of the coherent state
decreases and the photon statistics become sub-Poissonian.

\subsection{Vertical polarization mode}

We will calculate the photon number in the vertical polarization mode,
$\bracket{a^\dag\sub{V} a\sub{V}}$, and the second-order correlation
function $\bracket{a^{\dag 2}\sub{V} a^2\sub{V}}$.
They are approximated as
\begin{align}
 \bracket{a^\dag\sub{V} a\sub{V}}
  = \bracket{a^{\dag 2}\sub{V} a^2\sub{V}} = \frac{( |\alpha |^2 T )^2}{8},
\end{align}
which shows that the number of the vertically polarized photons,
as well as the second-order correlation function, grows in 
proportional to the second-order correlation function 
of the horizontal polarization mode.
The normalized second-order correlation function is obtained as
\begin{align}
 g^{(2)}(0) = \frac{\bracket{a^{\dag 2}\sub{V} a^2\sub{V}}}
{\bracket{a^\dag\sub{V} a\sub{V}}^2} = \frac{8}{(|\alpha|^2 T)^2} \gg 1.
\end{align}
Owing to the pair creation of photons, the strongly bunched photons appear
in the vertical polarization mode.

Similar calculations lead to
$\bracket{a\sub{V}}=0$ and $\bracket{a^2\sub{V}} = - \alpha^2 T/2$.
The evolution of $\bracket{a^2\sub{V}}$ is sensitive to the phase of
the coherence of the horizontal polarization mode.
The dispersion $\bracket{(\Delta X\sub{V}(\phi))^2}$
of an Hermitian operator,
$X\sub{V}(\phi)=a\sub{V} e^{- i \phi} + a^\dag\sub{V} e^{i \phi}$,
is approximated as
\begin{align}
 \bracket{(\Delta X\sub{V}(\phi))^2} =  1 - |\alpha|^2 T \cos 2({\rm arg} 
 \, \alpha-\phi), \label{Eq:Xb}
\end{align}
which is the same as Eq.~(\ref{Eq:Xa}).
The state of the vertically polarized photons is determined
by that of the horizontally polarized photons
and the directions of squeezing are the same. 
Figure \ref{Fig3} (b) illustrates the evolution 
from the vacuum state to the squeezed vacuum in the phase space.
Unlike the coherent squeezed state in Fig.~\ref{Fig3} (a),
the squeezed vacuum exhibits super-Poissonian statistics \cite{Walls:BOOK}.

\section{Conclusion and Discussion}

We have proposed the unique method of generating photon pairs 
with the polarization-dependent two-photon absorption.
The results of the perturbative calculations with the master equation
show that the vertically polarized photons and 
its second-order correlation function are increased together with 
the second-order correlation function of the horizontally polarization mode.
Moreover, the phase space representation
shows that the coherent state in the
horizontal polarization mode is changed into antibunching state,
and the vacuum state in the vertical polarization mode is developed into
a squeezed vacuum whose photon statistics is bunched.
The photon bunching is caused by the pair creation.

The generation of photon pairs at frequency $\omega_0$
by parametric down-conversion 
requires a pump beam at frequency $2\omega_0$,
so it is difficult to obtain photon pairs in the blue or shorter
wavelength region. 
On the other hand,
all photons in our method have the same frequency $\omega_0$.
By using appropriate two-photon absorbers,
we could produce photon pairs in blue or higher frequency range.

We estimate the possibility of experimental implementation
when we use a CW laser.
Using the two-photon absorption coefficient $\beta$,
$\kappa$ introduced in Sec.~III is expressed as
$\kappa=\beta \hbar \omega_0 c/(n^2 AL)$,
where $n$, $A$, and $L$ are the refraction index, 
the laser beam cross section, and the interaction length, respectively.
For the 6S-8S two-photon transitions of cesium at $822 \, {\rm nm}$,
the two photon absorption coefficient is $\beta \sim 1.8
\times 10^{-11} \,{\rm m/W}$
for atomic vapor at room temperatures.
When setting $L \sim 2 \times 10^{-2} \, {\rm m}$ and the laser
intensity $I \sim 1 \times 10^{10} \, {\rm W/m^2}$, and assuming
$n \sim 1$,
one obtains $N\sub{pair} = c \beta^2 I^2 L /(8n) \sim 2.4 \times 10^4$
photon pairs per second.
Although $N\sub{pair}$ seems to be a moderate value,
the experiment under this condition may be hard to carry out.
It is difficult to separate photon pairs from 
intense laser beam with orthogonal polarization.
For the laser beam cross section $A \sim 10^{-10} \,{\rm m^2}$,
the ratio between $N\sub{pair}$ and the photon number
of the laser beam per second $N\sub{laser} = A I/(\hbar \omega_0)$
is given as $R \equiv N\sub{pair}/N\sub{laser} \sim 5.6 \times
10^{-15}$,
but the extinction ratio of commercially available
linear polarizers is $10^{-6}$ at best.
The ratio $R$ can be improved by increasing the atomic density
or confining the vapor in a cavity.

It is known that the two-photon transition for the $\Gamma_1$
biexciton in CuCl has the same selection rule \cite{Tomita:PLA01}.
It has the advantage of large two-photon absorption coefficient
$\beta \sim 10^{-3} \, {\rm m/W}$.
Other parameters are $n \sim 3$ and $ \hbar \omega_0=3.2 \, {\rm eV}$. 
If we choose $I \sim 10^{4} \,{\rm W/m^2}$ and 
set the other parameters the same as in the case of Cs,
we get $N\sub{pair} \sim 2.5 \times 10^7 {\rm /s}$ 
and $R \sim 1.2 \times 10^{-5}$, which is large enough to discriminate.
While we have assumed the use of CW lasers for estimation,
the use of a pulsed laser would certainly ease the experimental
conditions.

In this paper we have dealt with the degenerate two-photon transition,
but it is easy to extend our method to a non-degenerate transition.
In the degenerate two-photon process,
two photons appear simultaneously in pair in the same mode,
while in the non-degenerate process
the pair of photons will emerge simultaneously but in different
frequency modes.
In the non-degenerate case,
it is possible to adjust the detuning $\Delta$ from the intermediate
level and we can easily attain large two-photon absorption coefficient
$\beta$,
which is inversely proportional to $\Delta^2$.
This strategy might enable us to demonstrate experiments
using atomic vapor.

\begin{acknowledgements}
This research was supported by the Ministry of
Education, Culture, Sports, Science and Technology
in Japan under a Grant-in-Aid for Scientific Research
No.~11216203 and COE Project No.~14213.
\end{acknowledgements}

\end{document}